\documentclass{aa}
\usepackage{graphics}
\begin{document}
\thesaurus{11(11.05.1; 11.05.02; 11.06.1; 13.09.1; 12.03.3;)}
\authorrunning{T. Treu et al.\ }
\titlerunning{An Extremely Red $r^{1/4}\,$ Galaxy}
\title{An Extremely Red $r^{1/4}\,$ Galaxy in the Test Image of the
Hubble Deep Field South\thanks{Based on observations collected with
the NASA/ESA HST, obtained at the STScI, which is
operated by AURA, under NASA contract NAS5-26555 and at the CTIO,
NOAO, which is operated by AURA, under cooperative agreement with the
NSF.}}

\author{T. Treu \inst{1,2} 
\and M. Stiavelli \inst{1,2,6}
\and A.~R. Walker  \inst{3}
\and R.~E.Williams \inst{2}
\and S.~A. Baum \inst{2}
\and G. Bernstein \inst{4}
\and B.~S. Blacker \inst{2}
\and C.~M. Carollo \inst{5}\thanks{Hubble Fellow}
\and S. Casertano \inst{2,6}
\and M.~E. Dickinson  \inst{2}
\and D.~F. Demello  \inst{2}
\and H.~C. Ferguson  \inst{2}
\and A.~S. Fruchter  \inst{2}
\and R.~A. Lucas \inst{2}
\and J. Mackenty  \inst{2}
\and P. Madau \inst{2} 
\and M. Postman \inst{2}}

\offprints{T. Treu}
\mail{T. Treu} 

\institute{Scuola Normale Superiore, Piazza dei Cavalieri 7, I56126,
Pisa, Italy; treu@cibs.sns.it, mstiavel@astro.sns.it
\and Space Telescope Science Institute, 3700 San Martin Dr., 21218 MD,
U.S.A.; treu@stsci.edu, mstiavel@stsci.edu, wms@stsci.edu
\and Cerro Tololo Inter-American Observatory, NOAO, Casilla 603, La
Serena, Chile; awalker@noao.edu \and University of Michigan, Dept. of
Astronomy, 830 Dennison Building, Ann Arbor, MI 48109, U.S.A.
\and Johns Hopkins University, 3701 San Martin Dr., 21218 MD, U.S.A.
\and On assignment from the Space Science Department of the European
Space Agency} \date{Received / Accepted}

\maketitle

\begin{abstract}
We report the serendipitous discovery of an extremely red object in
the Hubble Deep Field South (HDFS) Test NICMOS (Near Infrared Camera
and Multi Object Spectrograph) field of view.  The object is resolved
in the NICMOS image and has a light profile very well described by an
$r^{1/4}\,$ law with effective radius $r_{\textrm{e}}=0\farcs
20\pm0.05$ and H$_{\textrm{{\sc ab}}}=21.7\pm0.1$ magnitudes. In
contrast, the galaxy is undetected in the R and I band ground based
images taken at the CTIO 4 m Blanco Telescope, giving a lower limit to
the color of (R-H)$_{\textrm{{\sc ab}}}$$>3.9$ and
(I-H)$_{\textrm{{\sc ab}}}$$>3.5$ at the 95 \% confidence level.  The
colors of a range of synthetic galactic spectra are computed, showing
that the object is likely to be an ``old'' elliptical galaxy at
redshift $z\ga 1.7$. Alternatively the colors can be reproduced by an
``old'' elliptical galaxy at somewhat lower redshift ($\ga 1$) with
significant amount of dust, or by a younger galaxy at higher redshift.
This object represents a very interesting target for future VLT
observations.

\keywords{Galaxies: elliptical and lenticular, cD--Infrared:
galaxies--Galaxies: formation--cosmology: observations--early universe}

\end{abstract}

\section{Introduction}
\label{sec:intro}

The joint effort of the Hubble Space Telescope (HST) and large
ground based telescopes in the last few years has produced a real
breakthrough in our understanding of the history of formation and
evolution of galaxies. Deep photometric multicolor imaging such as the
Hubble Deep Field (HDF, Williams et al. 1996) together with
spectroscopic information (e.g., \cite{CFRS}; \cite{S96}; \cite{DEEP})
have started to sketch a sort of cosmic timetable that provides
important constraints to theories for the formation of cosmic
structure (\cite{S98}; \cite{CF}).  Besides global properties
(\cite{MP}), the superb resolution of the HST allows us to study the
evolution of different morphological types separately, in particular
to clarify the history of elliptical galaxies (E/S0). The intermediate
redshift spectroscopic data (e.g. \cite{DFKI98}) suggest a mostly
passive evolution of old stellar populations that seems to be
confirmed by photometric studies at higher redshift
(\cite{K97}; \cite{EE}).

To push the investigation to redshift significantly greater than 1,
infrared (IR) photometry (corresponding to optical rest-frame
emission) is needed (e.g. \cite{Mao}; \cite{Dick}). The extremely red
objects with (R$-$K)$>5$ (equivalently (R$-$K)$_{\textrm{\sc
ab}}>3.3$) found with IR surveys are generally thought to be high
redshift elliptical galaxies (e.g., \cite{Hu}; \cite{SP}). However,
due to their subarcsecond sizes, HST is needed to explore in detail
the morphology of these objects (\cite{GD}).

In this Letter we report the discovery of a ``R-H dropout''
(Figs.~\ref{fig:IM} {\bf a}, {\bf b} and {\bf c} in the HDFS test
image. The morphology, the size and the colors (R-H and I-H) strongly
suggest the object to be an ``old'' elliptical galaxy (or
bulge-dominated object) at $z\ga 1.7$ and therefore with a very high
formation redshift ($z_{\textrm{f}}$), somehow similar to the ``old''
red galaxy found by Spinrad et al. (1997) at redshift $z=1.55$.  Many
extremely red objects have been found in the last years (see
e.g. \cite{E}; \cite{DSD}; \cite{Mou}; \cite{Sta}) using ground based
IR photometry, thus with poor morphological information. At the
opposite, the object we present in this Letter, thanks to the HST
angular resolution, is resolved and we can measure the light profile
and the effective radius (similar objects are being detected with
NICMOS, e.g. Mc Carthy et al, 1998, but the luminosity profile has not
been measured). The luminosity profile is well described by an
$r^{1/4}\,$ law. Models with dust are also considered to explain the
extremely red colors, given the growing evidence for significant
amounts of dust in high redshift galaxies (\cite{So98};
\cite{Cima}). The Hubble constant is assumed to be 100 $h$ km/s/Mpc,
with $h=0.65$ where needed.

\section{Photometry}

\label{sec:photo}

During the HDFS test program (Williams et al. 1997) images of the
candidate HDFS field were taken with the F160W filter of the Near
Infrared Camera and Multi Object Spectrometer (NICMOS) on board the
HST (Fig.~\ref{fig:IM} {\bf a}), and in the optical R band
(Fig.~\ref{fig:IM} {\bf b}) with the 4m Blanco Telescope at the Cerro
Tololo Inter-American Observatory (CTIO). Follow up I band photometry
was obtained on May 13 1998 at the same telescope (Fig.~\ref{fig:IM}
{\bf c}).

\subsection{The Infrared NICMOS Imaging}

\label{ssec:NICMOS}

\begin{figure}
\resizebox{\hsize}{!}{\includegraphics{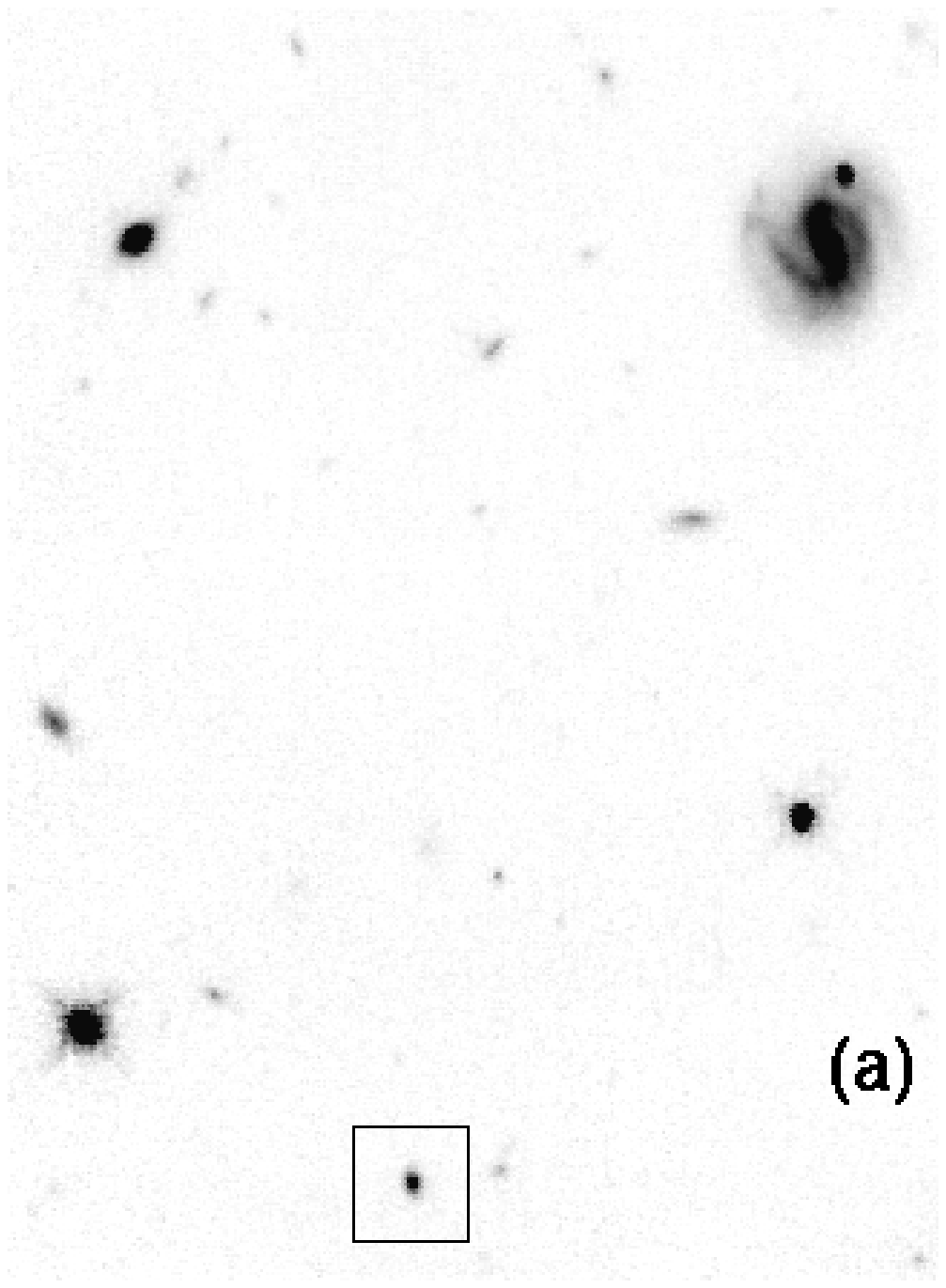}\includegraphics{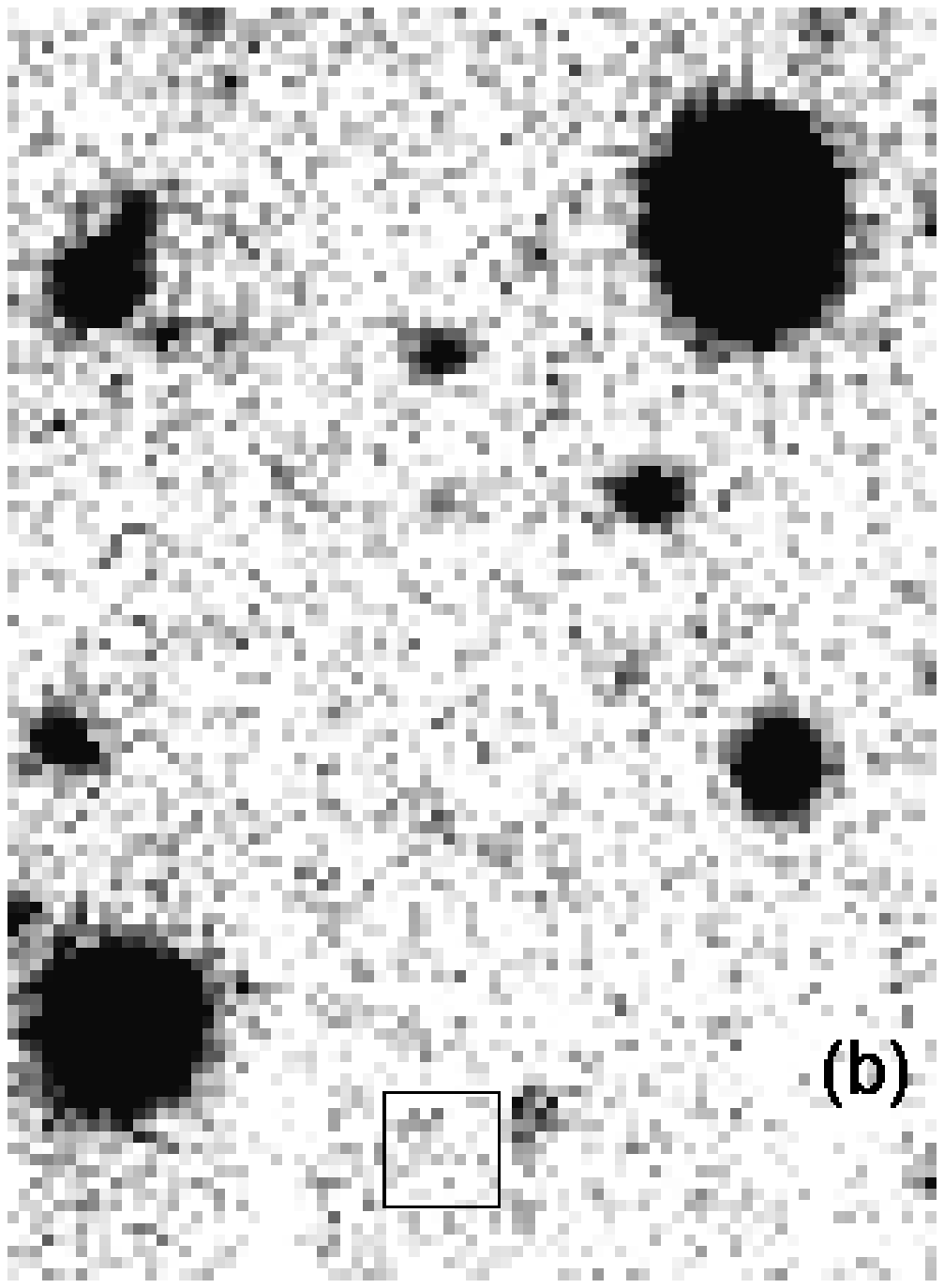}\includegraphics{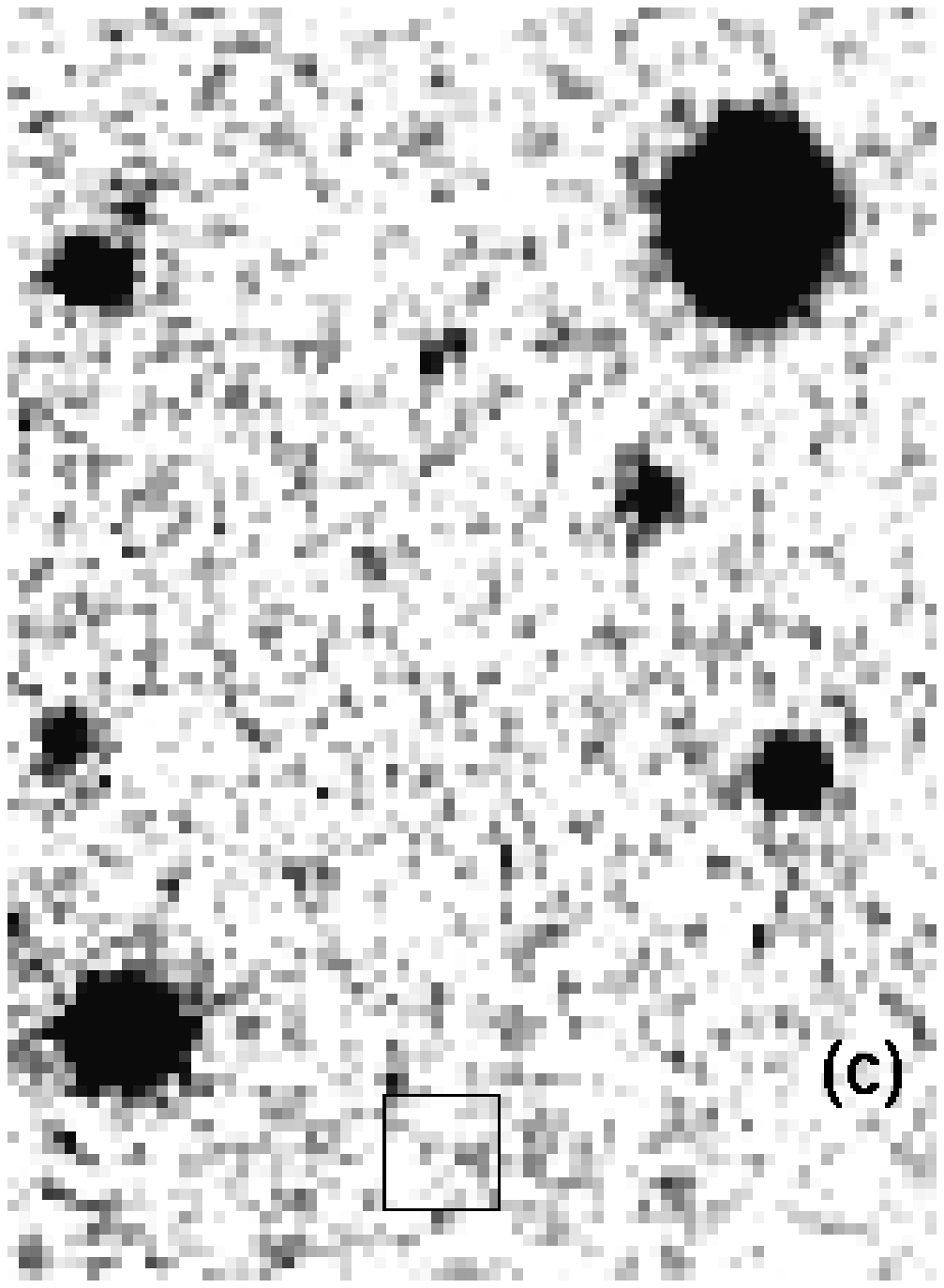}}
\caption{{\bf a.} HST NICMOS F160W ``drizzled'' image (7680 s). {\bf
b.} CTIO 4m R (3000 s). {\bf c.} CTIO 4m I (2800 s).  North is up and
East is left in the CTIO exposures, while the NICMOS image is slightly
rotated.  The frames are $\approx 35'' \times 45''$. The boxes are
centered on the position of the galaxy.}
\label{fig:IM}
\end{figure}

Twelve NICMOS Camera 3 (NIC3) images of the field were taken in
MULTIACCUM mode (640 s each). The images were obtained at six
different pointings in order to correct for bad pixels and to recover
the information lost to undersampling of NIC3 (approximately $0\farcs
2$ pixel size).

The ``pedestal'', an unpredictable bias that affects the NICMOS images
(see the NICMOS Image Anomalies web page at URL
www.stsci.edu/ftp/instrument\_news/~NICMOS/nicmos\_anomalies.html),
was removed using the Pedestal Estimation Software by R.~P. van
der Marel, and scripts developed by one of us (MED).  The dithered
images were combined on a subsampled grid ($0\farcs$1 pixel size,
shown in Fig.~\ref{fig:IM} {\bf a}) using the {\sc drizzle iraf/stsdas} task
(\cite{drizzle}). A bad pixel and cosmic ray mask was obtained by
using the {\sc iraf/stsdas} package {\sc dither}II.

The object appears resolved on the final drizzled images, but to be
sure that this was not an artifact of the reduction procedure we
performed a $\chi^2$ test on the single calibrated frames. For each
image we measured the FWHM along the x and y axis of the galaxy and
the two nearby stars (using the {\sc midas} command {\sc
center/gauss}). We estimated the error on the FWHM to be the standard
deviation of the values measured on the different frames (about
$0\farcs 04$). The reduced $\chi^2$ of the values with respect to
their average is therefore 1 (by definition).  The reduced $\chi^2$ of
the measured FWHMs computed with respect to the average stellar FWHM
is $\chi^2=11.1$, showing that the object is resolved.

\begin{figure}
\resizebox{\hsize}{!}{\includegraphics{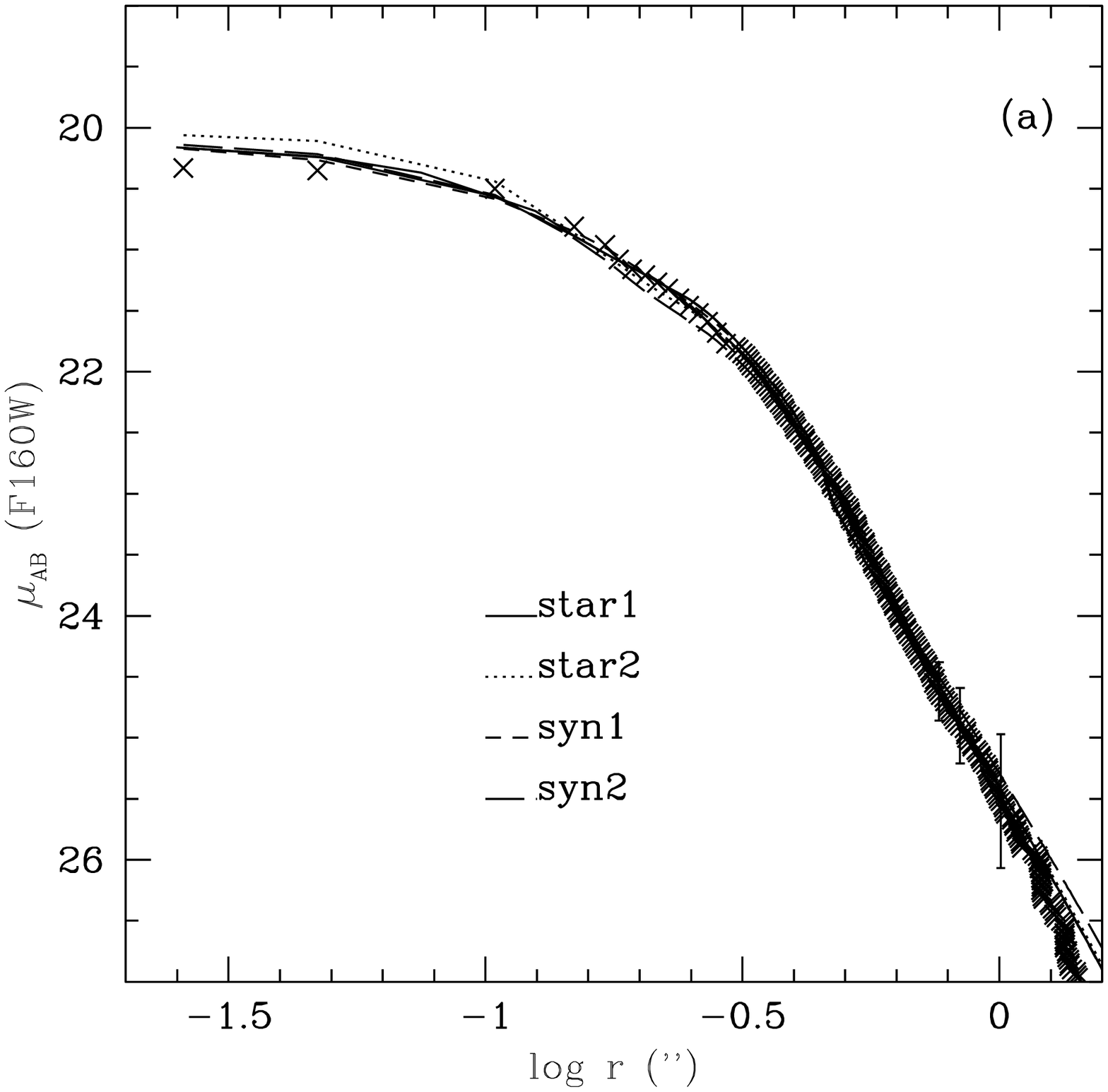}\includegraphics{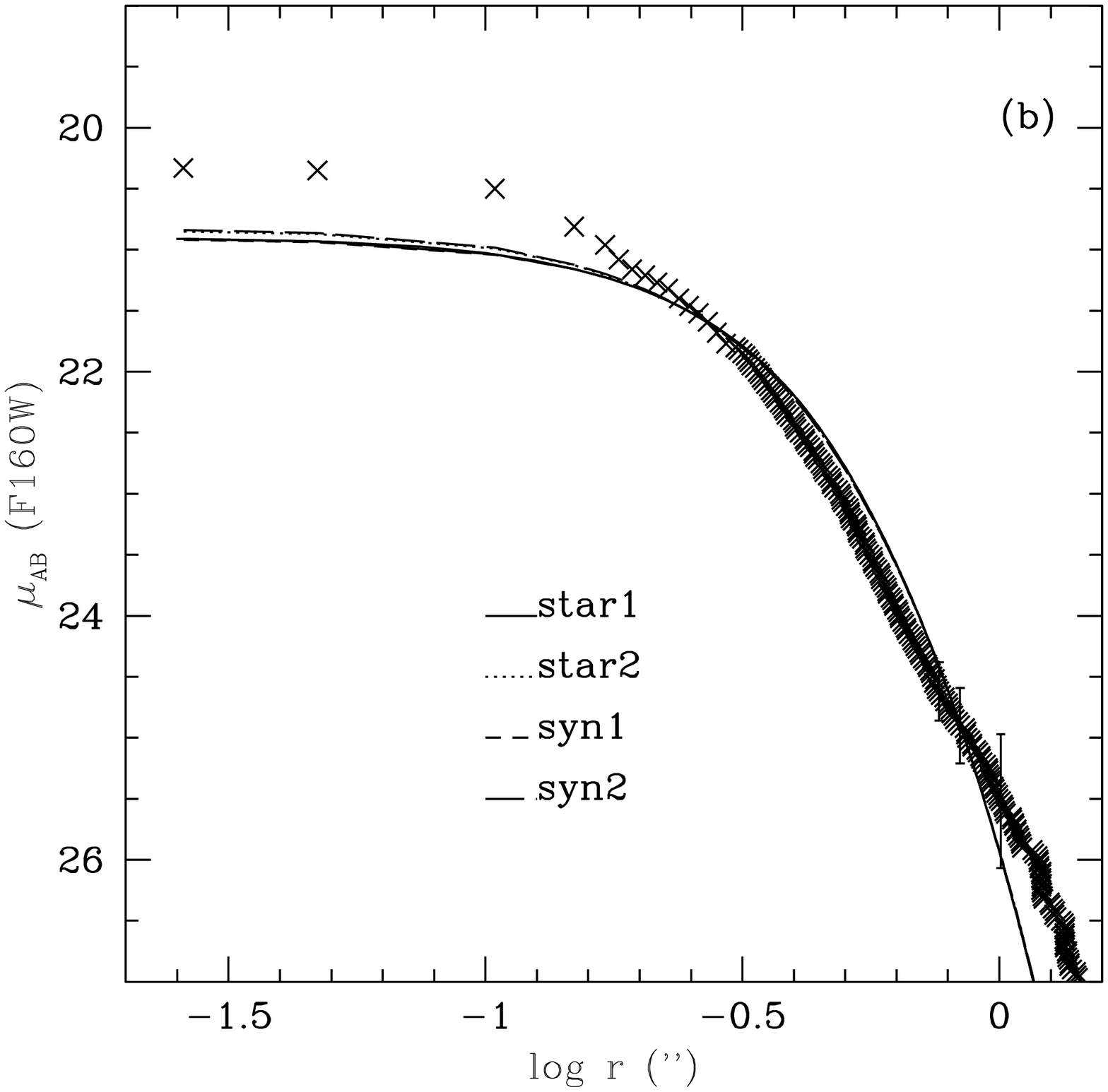}}
\caption{IR (F160W) luminosity profile of the galaxy (crosses). The
PSF-convolved best-fit $r^{1/4}\,$ ({\bf a}) and exponential ({\bf b})
profiles are shown. Only a few error bars are shown for clarity. Four
different fits are obtained by using different PSF profiles as
described in the text.}
\label{fig:prof}
\end{figure}

The luminosity profile of the galaxy (Fig.~\ref{fig:prof}) was
obtained by fitting elliptical isophotes to the image with a modified
version of the {\sc midas} command {\sc fit/ell3} specifically
designed to deal with undersampled images (M{\o}ller, Stiavelli \&
Zeilinger, 1995).

In order to fit a luminosity profile law to the data point it is
crucial to determine the PSF very accurately.  An unexpected
deformation of the NICMOS dewar pushed NIC3 out of the range reachable
by the focusing device (Pupil Alignment Mechanism) and therefore NIC3
is always somewhat out of focus.  Thermal breathing of the instrument
(that is significant during long exposures) and the combination of
multiple dithered exposures add further uncertainty and broadening to
the PSF. For these reasons the stars in the field are a better
approximation of the real PSF than the PSFs obtained with Tiny Tim
4.4; the main residual difference may be caused by position-dependent
features of the NIC3 PSF. In order to have an estimate of the errors
due to the PSF, we fitted the profile with different PSFs: the two
nearest stars and two ``synthetic'' ones computed as follows. A PSF
was calculated using Tiny Tim 4.4 (using a 15 mas jitter) on a
subsampled grid ($0\farcs02$ pixel size) in the same position of a
star in the HDFS test image. It was then rebinned to the scale of the
drizzled image in order to obtain a PSF centered in the same sub-pixel
of the real star. Under the assumption that the ``true'' PSF ($S$) is
the convolution of the Tiny Tim PSF ($T$) with a position independent
broadening function ($B$) due mostly to the breathing of NICMOS and
the drizzling of the single dithered images, we derived the function
$B$ using the {\sc midas} fourier transform commands. The broadening
functions $B_1$ and $B_2$ were derived for the two nearest
stars. Using a Tiny Tim PSF ($T_{\textrm g}$) centered on the galaxy
center and the broadening functions we obtained two ``synthetic
stars'' syn1=$T_{\textrm g}\circ B_1$ and syn2$=T_{\textrm g}\circ
B_2$.  As we were interested in checking the robustness of the results
with respect to PSF uncertainties, we performed the fits using both
the two real stars and the two ``synthetic'' stars.

We fitted an $r^{1/4}\,$ law to the isophotal profile, using a code
described, e.g., by Carollo et al.\ (1997). The fit obtained with the
four PSFs are in excellent mutual agreement and reproduce the data
very well (Fig.~\ref{fig:prof} {\bf a}). If an exponential profile
component is added, the best fit is obtained for a very small
exponential component and a practically unchanged $r^{1/4}\,$
component. At the opposite, an exponential disk by itself gives a poor
description of the profile (Fig.~\ref{fig:prof} {\bf b}). The
photometric parameters were also determined with a two-dimensional fit
code, described in Treu et al.\ (1998).  The total error is mainly due
to residual PSF mismatches, to the specific technique (profiles or 2D
fit), and to errors in the sky subtraction. The PSF and the technique
dependent uncertainties can be estimated from the scatter of all the
photometric parameters derived, while the sky-subtraction error has
been estimated by varying the sky level ($\approx 15$ \% in
$r_{\textrm{e}}$ and $\approx 0.1$ mag in m). The average values of
the effective radius and magnitude, obtained with different
techniques, models and PSFs, are respectively $r_{\textrm{e}}=0\farcs
20 \pm0.05$ and m=$21.7\pm0.1$.

\subsection{The R and I ground-based photometry}

\label{ssec:R}

Ten dithered R band exposures (300 s each) were obtained with the
Prime Focus CCD (PFCCD) Imager at the CTIO Blanco 4m telescope on May
2 1997, using the detector SITe 2K \#6.  The night was photometric
with seeing $1\farcs2$ . The same field was observed in the I band on
May 13 1998 (unfortunately almost Full Moon was present) with the
Tyson-Bernstein Mosaic Imager.  CCD \#3, also a SITe 2K, was pointed
at the PFCCD field, obtaining 14 x 200 s dithered exposures.  The
night was photometric, with $1\farcs0$ seeing.  The images were
reduced and combined in standard way using {\sc iraf}.

The photometric calibration of the R band was done by bootstrapping
the photometry from a 0.9m to the 4m image. Three relatively bright (I
$\sim$ 18) stars near the NICMOS position were calibrated on the
I-band frame by reference to seven stars in Landolt (1992) SA 107.
Both were also checked with the nominal zeropoints from the instrument
manuals. The accuracy of the calibration is about 0.1 mag, which is
good enough for our purposes.

The galaxy, which is quite evident on the NICMOS images, is undetected
in both the R and I CTIO image (Figs.~\ref{fig:IM} {\bf b} and {\bf
c}). Thus we can only set an upper limit to the R and I band
luminosities. The nearest detected source is at $\approx 3''$ on the R
band image.  The limiting magnitudes, computed by measuring the sky
noise ($\delta$K) and correcting the resulting flux for the seeing
losses, are m$_{\textrm{\sc r}}>25.9$ and m$_{\textrm{\sc i}}>$24.7 AB
mag at the two-sigma level.

A more accurate limit can be given by considering the statistical
distribution of the counts on the detector for a given source of
intensity $I_0$ and a given background of zero average and variance
$\delta$K$^2$.  The 95 \% confidence level (CL) limit is
m$_{\textrm{\sc r}}>25.6$ and m$_{\textrm{\sc i}}>$25.2 ABmag.  The
limits on the color are therefore (R-H)$_{\textrm{{\sc ab}}}$$>4.3$
and (I-H)$_{\textrm{{\sc ab}}}$$>$3, at the two-sigma level,
considering only the sky variance, but (R-H)$_{\textrm{{\sc
ab}}}$$>$3.9 and (I-H)$_{\textrm{{\sc ab}}}$$>$3.5, at the 95 \%
CL with the more accurate algorithm.

\section{Identification}

\label{sec:Id}

\begin{figure}
\resizebox{\hsize}{!}{\includegraphics{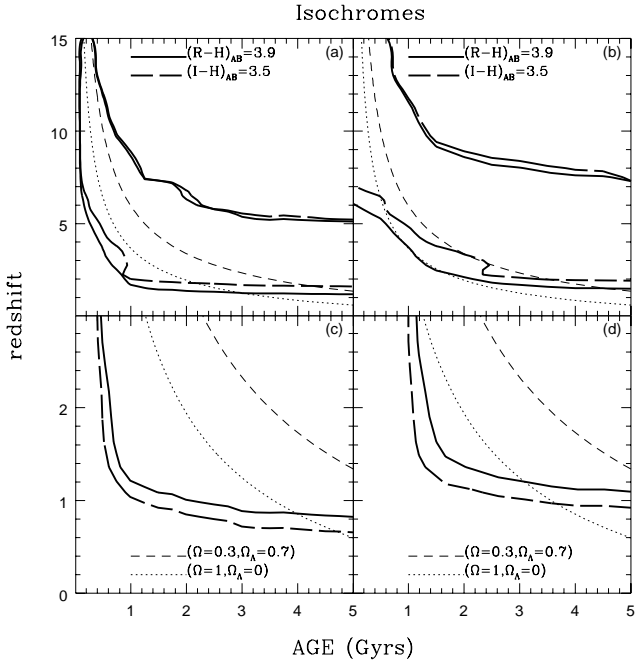}}
\caption{{\bf a.} Contour plot (thick lines) of the
(R-H)$_{\textrm{{\sc ab}}}$\, (solid) and (I-H)$_{\textrm{{\sc
ab}}}$\, (dashed) colors (isochrome) as a function of redshift and age
of an elliptical galaxy, computed using synthetic spectra of solar
metallicity.  The age of the Universe as a function of redshift is
overplotted as thin lines for two different cosmological models. See
text for discussion and details.  {\bf b.} As in {\bf a} but the
models have metallicity $Z=0.2 Z_{\odot}$.  {\bf c.} As in {\bf a} but
the spectra have been reddened for dust ($A_V=0.5$, see text).  {\bf
d.} As in {\bf c} but the models have metallicity $Z=0.2 Z_{\odot}$.}
\label{fig:za}
\end{figure}

The morphology and the IR light profile suggest that the
object is a high redshift ``old'' elliptical galaxy. But the strongest
clues in favor of a high redshift elliptical galaxy are the
colors. To better constrain this identification we have computed the
(R-H)$_{\textrm{{\sc ab}}}$\, and (I-H)$_{\textrm{{\sc ab}}}$\, colors
of a set of synthetic spectra of single burst elliptical galaxies
(Bruzual \& Charlot, 1993; GISSEL 96 version) as a function of
redshift. A standard Salpeter IMF and metallicities $Z=Z_{\odot}$ and
$Z=0.2 Z_{\odot}$ have been used.  The lower metallicity has been
selected to be representative of the low metallicity environment that
one could expect to find in primordial stellar populations.  In
Figs.~\ref{fig:za} {\bf a} and {\bf b} the contour plots of
(R-H)$_{\textrm{{\sc ab}}}$\, and (I-H)$_{\textrm{{\sc ab}}}$\,
(isochromes) as a function of redshift and age of the synthetic galaxy
are shown.  For each given redshift the galaxy age has to be less than
the age of the Universe, which is overplotted as thin lines for two
different values of the cosmological parameters $\Omega$\, and
$\Omega_{\Lambda}$.  In practice, only the points above the maximum of
the lower branches of the two isochromes (I and R), and below the thin
line for the selected cosmology, are consistent with the observed
colors. As can be seen, the colors support the identification
as an ``old'' elliptical galaxy at redshift $\ga 1.7$ for the solar
metallicity models. The lower limit in redshift is even more stringent
if we consider low metallicity spectra, because the Balmer Jump is
intrinsically shallower. A redshift $z\ga 1.7$ would also give an
effective radius in the usual range for elliptical galaxies ($\ga 1$
kpc). The colors are well reproduced also for much higher redshift
($\sim$7--15).

A significant content of dust (which can be found in high
redshift star-forming galaxies, see e.g. Soifer et al.\ 1998, Cimatti
et al.\ 1998) could redden the colors of a given galactic spectrum
and redshift. For this reason, the same isochromes were computed after
reddening the spectra for dust absorption following the extinction law given by
Cardelli et al.\ (1989) with A$_{\textrm{{\sc v}}}=0.5$,
A$_{\textrm{{\sc v}}}=3.1$E(B-V). In Figs.~\ref{fig:za} (c) and (d) the
isochromes are shown in the low redshift ($0$--$3$) range.  The colors
are now reproduced by lower redshift models ($z\ga 1$), and therefore
the object could be a high redshift ``old'' elliptical galaxy with
significant dust content. 

\section{Conclusions}

\label{sec:conc}

The object, that has been selected as an ``R-H dropout'' in the test
image of the HDFS, is clearly resolved on the NICMOS image and shows
an $r^{1/4}\,$ law profile with an effective radius of
$0\farcs20$. The colors (R-H)$_{\textrm{{\sc ab}}}$\, and
(I-H)$_{\textrm{{\sc ab}}}$\, are well reproduced by synthetic models
of ``old'' elliptical galaxies at redshift $z\ga 1.7$ or at higher
redshift (up to $\sim$ 15) by younger ones (Sect.~\ref{sec:Id}),
implying therefore a very high formation redshift. Alternatively it
could be a lower redshift ($z\ga 1$) elliptical galaxy with a
significant amount of dust. High redshift clusters of galaxies have
been photometrically identified in the vicinity of QSOs (\cite{z238};
\cite{S98}).  Therefore this galaxy may possibly be a companion to the
QSO J2233-60, which is located $\approx 8'$, i.e.  $\approx 2 h^{-1}$
Mpc (physical distance) from the HDFS NIC3 field. This identification
would then lead to $z\approx2.22$ compatible with the colors and size
that we have observed. At this redshift, H$_{\textrm{\sc ab}}$=21.7
would imply an absolute magnitude M$_{\textrm{\sc v}}=-22.7$
($\Omega$=1, $\Omega_{\Lambda}$=0) and M$_{\textrm{\sc v}}=-23.3$
($\Omega$=0.3, $\Omega_{\Lambda}$=0.7) which are normal for a 1.5
Gyr-old elliptical\footnote{A 1.5 Gyrs synthetic spectrum
(\cite{BC93}) of solar metallicity was used to compute the
K-correction and the filter transformation.}.

Further spectroscopic and photometric investigation are needed to
measure the redshift, to study in detail the stellar population, to
clarify the role of dust, and to identify possible high redshift
companions.  The faintness of the object though, makes it extremely
hard to measure the redshift with 4m class telescopes, while it
will be feasible in a few hours with the ISAAC spectrograph at the
VLT (see the VLT-ISAAC web page www.eso.org/instruments/isaac).

\begin{acknowledgements}

The authors would like to thank R. P. van der Marel, who
developed the Pedestal Estimation Software used for the pedestal
subtraction. TT would like to thank Prof. G. Bertin for carefully
reading the manuscript.

\label{sec:ak}

\end{acknowledgements}

\end{document}